\documentclass[twocolumn]{aastex631}

\usepackage{hyperref}
\usepackage{subfigure}
\usepackage{graphicx}
\usepackage[flushleft]{threeparttable}
\usepackage{amssymb}
\usepackage{soul}

\begin{document}

        \title{Accretion Geometry of the New Galactic Black Hole Candidate AT2019wey in the Hard State }

\correspondingauthor{Parijat Thakur}
\email{parijat@associates.iucaa.in, parijatthakur@yahoo.com}

\author[0009-0005-4510-4051]{Pragati Sahu}
\affiliation{Department of Pure and Applied Physics, Guru Ghasidas Vishwavidyalaya (A Central University), Bilaspur (C. G.)—495009, India}

\author[0000-0003-3499-9273]{Swadesh Chand}
\affiliation{Inter-University Centre for Astronomy and Astrophysics, Pune, Maharashtra-411007, India}
\affiliation{Institute of Astronomy, National Tsing Hua University, Hsinchu 300044, Taiwan}
\author[0000-0003-1589-2075]{Gulab C. Dewangan}
\affiliation{Inter-University Centre for Astronomy and Astrophysics, Pune, Maharashtra-411007, India}

\author[0000-0002-0333-2452]{Andrzej A. Zdziarski}
\affiliation{Nicolaus Copernicus Astronomical Center, Polish Academy of Sciences, Bartycka 18, PL-00-716 Warszawa, Poland}

\author{Vivek K. Agrawal}
\affiliation{Space Astronomy Group, ISITE Campus, U. R. Rao Satellite Center, Outer Ring Road, Marathahalli, Bangalore—560037, India}

\author[0000-0001-5958-0562]{Parijat Thakur}
\affiliation{Department of Pure and Applied Physics, Guru Ghasidas Vishwavidyalaya (A Central University), Bilaspur (C. G.)—495009, India}








\begin{abstract}
We perform broadband spectral and timing studies of the Galactic low-mass black hole candidate AT2019wey using quasi-simultaneous NICER, Swift, and NuSTAR observations obtained in 2022. The long-term MAXI light curve, along with the hardness–intensity diagram (HID), indicates that the source remained in the hard state and did not switch to the soft state. Spectral modeling using two different model combinations reveals that the broadband spectrum is best described by two distinct Comptonizing regions, associated reflection components, and thermal emission from the disk. The harder Comptonizing region dominates ($\gtrsim80\%$) the total flux and is primarily responsible for the observed reflection features from the distant part of the disk. We find that the accretion disk is truncated at a radius of $\sim16-56~r_{\rm{g}}$, while the luminosity is $\sim1.9\%$ of the Eddington limit, assuming a black hole mass of $10 ~ M_\odot$ and distance of 8 kpc. Our spectral results also show consistency in the estimated inner disk radius obtained through two independent methods: modeling the disk continuum and the reflection spectrum. The variability studies imply the presence of intrinsic disk variability, likely originating from an instability in the disk. We also detect hard time lags at low frequencies, possibly arising from the inward propagation of mass accretion rate fluctuations from the outer to the inner regions of the accretion disk. Moreover, an observed deviation of the lag-energy spectrum from the log-linear trend at $\lesssim 0.7$ keV is most likely attributed to thermal reverberation, arising from the reprocessing of hard coronal photons in the accretion disk.
\end{abstract}



\section{Introduction} \label{sec:intro}

Black hole X-ray binaries (BHXRBs) are powered by accretion of matter from the companion star onto the black hole. The emitted spectrum from these systems typically comprises three main components: thermal emission from the accretion disk, a non-thermal power-law component produced via Comptonization of soft disk photons by high-energy electrons in a hot corona, and a reflection component resulting from the reprocessing of high-energy photons by the accretion disk. The characteristics and relative contributions of these spectral components on the overall flux evolve as the system undergoes spectral state transitions during an outburst. The primary spectral states that the BHXRBs live through during a typical outburst are the low/hard state (LHS) and the high/soft state (HSS). In addition, two short-lived intermediate states, namely, the hard and soft intermediate states (HIMS and SIMS), are present between the LHS and HSS. All these spectral states and their associated timing properties can be effectively distinguished through a q-shaped diagram, commonly known as the hardness-intensity diagram \citep[HID;][]{Belloni2005}.

Understanding the geometry of BHXRBs is crucial for probing the underlying accretion mechanisms. Despite decades of observational and theoretical studies, the detailed structure and geometry of the accretion disk and the corona in these systems remain unclear. It is now widely accepted that in the HSS, the accretion flow is dominated by thermal emission peaking at $\sim1$ keV, originating from an optically thick and geometrically thin disk that extends down to the innermost stable circular orbit \citep[ISCO;][]{1973A&A....24..337S, Gierlinski2004, McClintock2006csxs.book..157M}. In this state, the source exhibits very low variability, typically $\lesssim 1\%$. On the other hand, the structure of the accretion disk and corona in the LHS remains a subject of ongoing investigation. The most widely accepted model to explain the disk–corona geometry in the LHS is that the accretion disk is truncated at a large radius from the ISCO, and the inner region is replaced by a hot, optically thin, and radiatively inefficient flow called as advection-dominated accretion flow \citep[ADAF;][]{Esin1997, Done2007}. Thermal Comptonization in the corona mainly governs the shape of the spectrum, and a large variability up to $\sim30\%-40\%$ on short timescales is observed in this state. However, the truncated disk scenario in the LHS has been challenged by the detection of broad iron K$\alpha$ lines, suggesting that the accretion disk may instead extend close to the ISCO in this state \citep{Miller2006ApJ...653..525M}. Additionally, several studies have reported that in the bright hard state, when the source luminosity exceeds $\sim1\%$ of the Eddington luminosity, the disk can extend close to the ISCO \citep{Reis2008MNRAS.387.1489R, Garcia2015ApJ...813...84G, Garcia2019ApJ...885...48G}. 

To reconcile this controversy, models with two separate Comptonizing regions have been proposed for the LHS rather than considering a single Comptonizing region for a homogeneous corona. The assumption of the dual-corona scenario has proven to be successful in explaining the truncated disk model, even in the bright LHS of BHXRBs. In this framework, the hard Comptonizing region is primarily responsible for the origin of the reflection features from the distant parts of the accretion disk, and the broadened reflection features arise due to the soft Comptonizing region \citep{Zdziarski2021ApJ...909L...9Z, Zdziarski2022ApJ...928...11Z, Banerjee2024ApJ...964..189B, Chand2024ApJ...972...20C}.

The strong X-ray variability observed in the LHS of BHXRBs can be associated with various physical processes and interactions among them. Studying this variability across different timescales offers an independent probe of the geometry of the accretion flow. The most common approaches to understand the nature and origin of the variability are the study of power density spectrum (PDS), fractional rms variability amplitude as a function of photon energy (rms-energy spectrum), and time lags as a function of both frequency and energy. In the LHS, the PDS is typically dominated by aperiodic broadband noise, often referred to as band-limited noise (BLN), and is sometimes accompanied by low-frequency quasi-periodic oscillations \citep[QPOs; see][for further details]{Belloni2002ApJ...572..392B, Casella2004A&A...426..587C}. Investigating these features provides crucial insights into the structure and dynamics of the accretion flow. Furthermore, studies of the rms-energy spectrum acts as a tool to identify energy dependence of variability by separating the variable component from the constant ones in the X-ray spectrum. Another key aspect to understand the nature of variability involves examining lag-frequency and lag-energy spectra. Time lags observed between different energy bands can be either positive or negative. A positive lag, commonly termed as hard lags, indicates that hard photons lag behind soft ones. Such hard lags are typically observed at low frequencies and are often interpreted as a result of the inward propagation of mass accretion rate fluctuations through the disk \citep{Lyubarskii1997, Arevalo2006MNRAS.367..801A}. Conversely, the soft lags, also called negative lags, are usually observed at relatively high frequencies, where soft photons lag behind hard ones. These soft lags are often interpreted as reverberation lags, arising from the light travel time delay between the primary continuum and reprocessed emission off the accretion disk \citep{Uttley2011MNRAS.414L..60U, DeMarco2015ApJ...814...50D}. Mapping reverberation lags thus plays a pivotal role in constraining the disk–corona geometry of BHXRBs.

AT2019wey (also known as SRGA~J043520.9$+$552226, SRGE\hspace{0pt}~J043523.3$+$552234) was first discovered as an optical transient in December 2019 by the Zwicky Transient Facility \citep[ZTF;][]{2019Masci} survey. The event was subsequently observed independently by the Asteroid Terrestrial-impact Last Alert System \citep[ATLAS;][]{2018PASP..130f4505T} and Global Astrometric Interferometer for Astrophysics \citep[Gaia;][]{2016gaia} surveys. Later, it was identified as an X-ray source by the extended ROentgen Survey with an Imaging Telescope Array \citep[eROSITA;][]{Predehl2021} and the Astronomical Roentgen Telescope - X-ray Concentrator \citep[ART-XC;][]{Pavlinsky2021} instruments onboard the Spectrum-Roentgen-Gamma (SRG) observatory in March 2020 \citep{2020ATelMereminskiy}. Initially, the source was classified as a supernova based on Gaia observations, and was later considered a potential BL Lac object \citep{2020ATel13576....1L}. However, follow-up observations by \citet{2020ATel13932....1Y}, which revealed hydrogen emission lines at zero redshift, suggested that AT2019wey is a Galactic low-mass X-ray binary (LMXB) hosting either a black hole or a neutron star as an accretor. Later on \cite{Yao_2021_ApJ_920_120} carried out a multiwavelength study of AT2019wey using observations in 2020 during the same outburst and concluded that the companion star is likely a low-mass ($<1~M_\odot$) object in a short-period orbit ($\lesssim 16~\mathrm{hr}$). The authors also found that the source location on the radio and optical luminosity versus X-ray luminosity diagrams aligns more closely with BHXRBs than with neutron star systems, further supporting its classification as a candidate BHXRB. Using optical and X-ray observations in 2020, \citet{Mereminskiy2022A&A...661A..32M} also proposed that AT2019wey is a low-mass BHXRB. In addition, a comprehensive X-ray analysis by \citet{Yao_2021_ApJ_920_121} during the 2020 outburst showed that the spectral and timing properties of AT2019wey are consistent with those of canonical BHXRBs. Further support came from a spectral study by \citet{Feng2022SCPMA..6519512F} using NuSTAR data, which reinforced the interpretation of the source as a BHXRB. It is noteworthy that the spectral modeling in all these previous X-ray studies of this source was performed using a single Comptonizing region, which could not constrain the electron temperature or the cut-off energy. Moreover, these studies reported a super-solar iron abundance and suggested that AT2019wey is a low-inclination system hosting a maximally rotating black hole.

\begin{table*}
	\caption{Observation details of 2022 outburst of AT2019wey}
   \centering
   \renewcommand{\arraystretch}{1.2}
   \begin{tabular}{cccccc}
      \hline
      \hline
  
    Observation  & Instrument & Observation ID & Start time & Stop time &  Eff. Exp. (s) \\
     \hline
     \hline
     Obs1 &Swift/XRT & 00089577001 & 2022-12-16 00:31:07 & 2022-12-16 00:46:56 & 931.5 \\
          & NuSTAR/FPMA & 80801328002 & 2022-12-15 12:11:09 & 2022-12-16 02:31:09  & 23810 \\ 
          & FPMB        & \nodata     & 2022-12-15 12:11:09 & 2022-12-16 02:31:09  & 23730 \\ \\

    Obs2 
     & NICER & 5701010104 & 2022-12-21 23:47:43 & 2022-12-22 09:29:00 & 8944\\ 
     & NuSTAR/FPMA & 80801328004 & 2022-12-21 19:26:09 & 2022-12-22 09:51:09  & 24030 \\
     & FPMB         & \nodata   & 2022-12-21 19:26:09 & 20200-12-22 09:51:09 & 23950 \\ \\
  \hline  
   \end{tabular}
   \label{tab:obs_ids}
\end{table*}

\begin{figure}
    \includegraphics[width=\columnwidth]{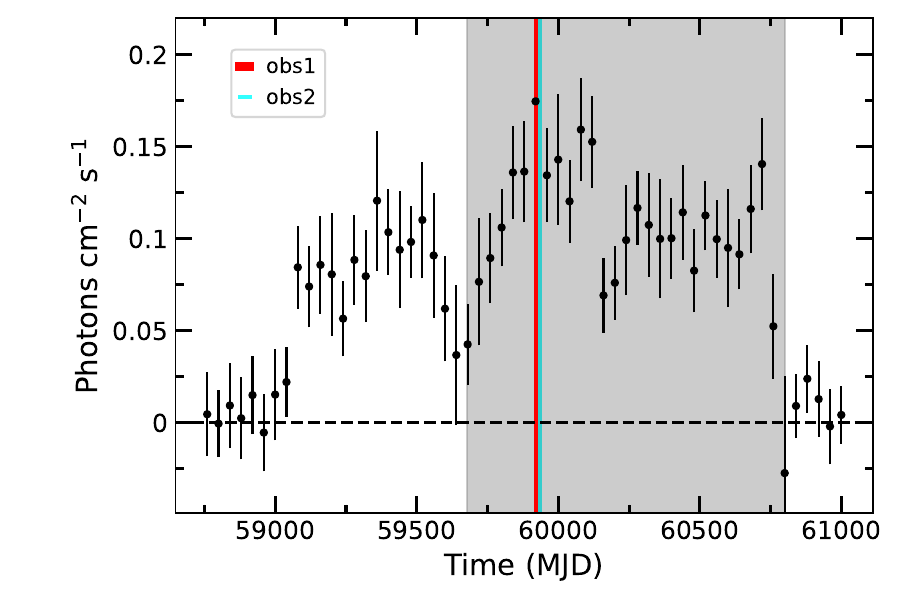}
    \caption{Long-term MAXI lightcurve of AT2019wey in the $2-20$ keV band binned with 40 days, showing the entire duration of the source activity including the onset of the first outburst, a brief quiescent period, and resumption in early 2022 lasting for another $\sim 3$ years.  The vertical lines indicate the timings of the observations used in this work. The shaded region marks the period from the onset to the decay of the outburst, during which the HID is derived (see Figure~\ref{fig:hid_nicer}).}
    \label{fig:maxi_ltcrv}
\end{figure}

In this paper, we present a detailed spectral and timing analysis of AT2019wey using quasi-simultaneous observations from NICER, Swift, and NuSTAR during its 2022 outburst in the hard state. Our broadband ($0.5-78$ keV) spectral modeling indicates the requirement of two distinct Comptonizing regions, rather than a single homogeneous corona, to successfully constrain the coronal temperatures. Our comprehensive timing analysis, including the study of PDSs, rms–energy spectra, and time lags as functions of both frequency and energy, provides valuable insights into the origin and nature of variability in the source. The structure of this paper is as follows. In Section 2, we describe the observations and data reduction procedures. Section 3 presents our spectral and timing analysis results. Finally, Section 4 is devoted to discussion and concluding remarks.

\begin{figure}
    \includegraphics[width=\columnwidth]{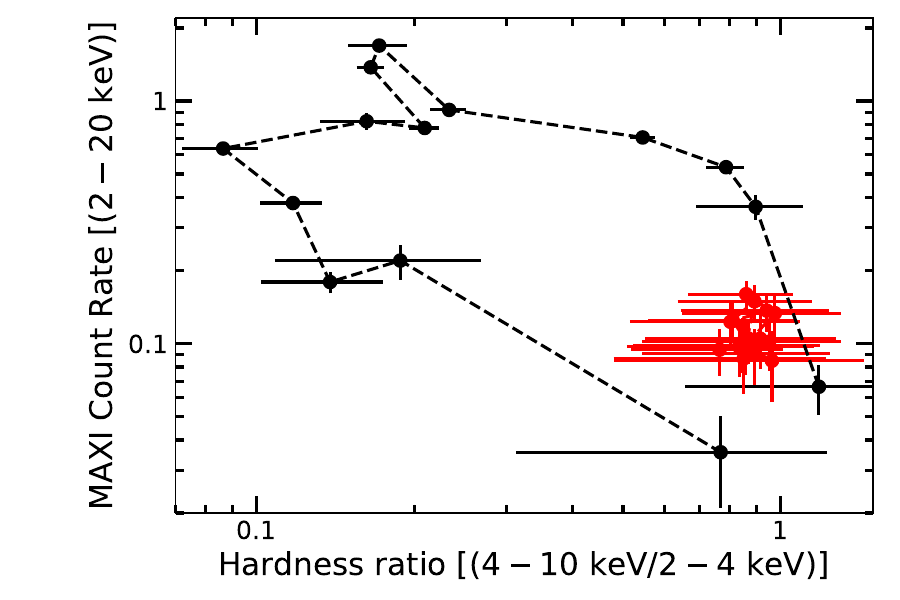}
    \caption{HID of AT2019wey constructed using MAXI observations (red) taken during the shaded interval shown in the MAXI light curve in Figure~\ref{fig:maxi_ltcrv}. The MAXI data have been binned for clarity. Black circles show the `q' - shaped HID of GX~339--4, derived from MAXI observations covering its full or successful outburst in 2021.}
    \label{fig:hid_nicer}
\end{figure}

\section{Observations} \label{sec:style}
The Galactic BHXRB AT2019wey went into its first detected outburst in 2019, which appears to have lasted until early 2022. Figure~\ref{fig:maxi_ltcrv} shows the long-term 40-day binned MAXI light curve of AT2019wey in the 2–20 keV band. After a brief quiescent period of about a month, the source resumed its outburst that continued until early 2025, which we refer to here as the ``2022 outburst". In this study, we analyze two X-ray observations of the source during the 2022 outburst, performed quasi-simultaneously by the Swift/X-ray telescope (XRT), NICER, and NuSTAR. The start dates of these observations are marked on the MAXI light curve in Figure~\ref{fig:maxi_ltcrv}, with detailed information provided in Table~\ref{tab:obs_ids}. We also utilized MAXI data from the gray shaded region in Figure~\ref{fig:maxi_ltcrv}, covering the full outburst, to construct the HID. The hardness ratio (HR) was calculated as the ratio of the $4-10$ keV to $2-4$ keV count rates, and was plotted against the $2-20$ keV count rate. The resulting HID is shown in Figure~\ref{fig:hid_nicer}. For comparison, we overplotted the full `q'-shaped HID of GX~339--4, derived from MAXI observations during its complete outburst in 2021. The comparison indicates that the 2022 outburst of AT2019wey was a failed, or hard-only, outburst, with the source remaining in the LHS throughout the entire outburst period.

\subsection{NuSTAR} 
NuSTAR \citep{Harrison2013ApJ...770..103H} observed AT2019wey twice during its outburst started in 2022. To analyze the NuSTAR Data, we used the standard analysis software \texttt{NuSTARDAS} (v2.1.2) along with the latest calibration database (CALDB version: 20230613) for both the FPMA and FPMB modules. The standard script \texttt{nupipeline} was employed to process the data. We then derived the source and background spectra using the \texttt{nuproduct} task, using circular regions of $60\arcsec$ radius centered at the source and away from the source, respectively. The corresponding RMFs and ARFs were also generated using the same task.

\subsection{Swift/XRT}
Swift observed AT2019wey once quasi-simultaneously with NuSTAR. We utilized this dataset obtained in the Windowed Timing (WT) mode by Swift/XRT \citep{Burrows2000SPIE.4140...64B, Hill2000SPIE.4140...87H}. The level1 data were processed to produce a level2 event file using the standard task \texttt{xrtpipeline} (v.0.13.7) and the latest calibration files. We then employed \texttt{xselect} (v.2.5b) to extract the source and background spectra, selecting circular regions with a radius of $30\arcsec$, centered at the source position and a region away from the source, respectively. The data were found to be free from pile-up, as the source count rate was well below the threshold value for pile-up in the WT mode. Subsequently, the ARF was generated using the standard task \texttt{xrtmkarf}. Source and background light curves were also extracted using the same circular regions, and the source light curve was then corrected for the background using the \texttt{lcmath} task.

\subsection{NICER}
NICER \citep{Gendreau2016SPIE.9905E..1HG} observed AT2019wey multiple times during its second outburst that started in 2022. Here, we have used one single NICER observation taken quasi-simultaneously with NuSTAR (see Table~\ref{tab:obs_ids}) for detailed spectral and timing studies. For the NICER data analysis, we employed the standard pipeline \texttt{NICERDAS}, available within HEASoft (v.6.33.2), along with the latest calibration database (CALDB version: xti20240216). We generated cleaned event file using the standard task \texttt{nicerl2}. We then used the standard script \texttt{nicerl3-spect} to extract the source and background spectra, where the \texttt{`3C50'} model was used to estimate the background. During the standard analysis, the {\texttt{nicerl3-spect} tool automatically incorporates a systematic of approximately $1.5\%$ to each channel in the $0.4-10$ keV band. The task \texttt{nicerl3-spect} also generates the corresponding response matrix file (RMF) and ancillary response file (ARF). Lightcurves were generated using the standard task \texttt{nicerl3-lc}.

\begin{figure}
    \includegraphics[width=\columnwidth]{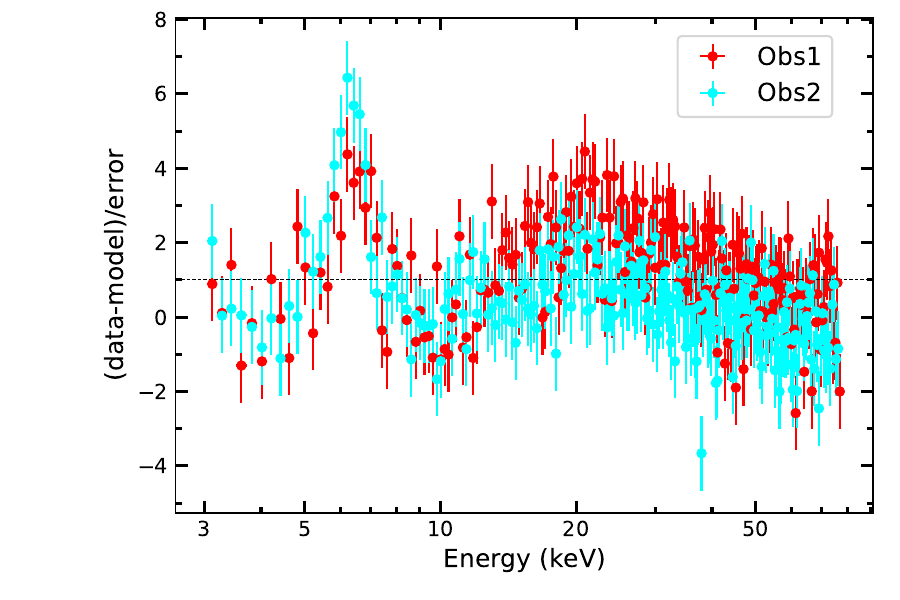}
    \caption{Residual of NuSTAR FPMA spectral data in $3-80$ keV to the fit with \texttt{tbfeo(powerlaw+diskbb)}. Spectral fitting is performed 
    only in the $3-5$ and $9-12$ keV bands. Presence of the iron line excess and reflection hump are clearly visible in both the dataset.}
    \label{fig:fpma_res}
\end{figure}

\begin{table*}
\caption{Best-fit broadband X-ray spectral parameters derived using Model-1}
\renewcommand{\arraystretch}{1.2}
\centering
 \begin{tabular}{lccc}
       \hline
      \hline
Component & Parameter & Obs1 & Obs2 \\
	\hline
 TBfeo & $N_H$ $(\times 10^{22}$ cm$^{-2}$) & $0.59^{+0.10}_{-0.05}$ & $0.62^{+0.02}_{-0.03}$ \\
            & $O$ & $1^f$ &  $1.0\pm0.1$  \\
            & $Fe$  & $0.5^f$ & $0.50^{+0.07}_{}$ \\ \\
	
Diskbb & $kT_\mathrm{in}$ (keV) & $0.2^f$ & $0.21^{+0.02}_{-0.01}$  \\
	  & norm &  $2147^{+5975}_{}$  & $11385.2^{+7570.4}_{-5273.5}$\\ \\
  
ireflect & $R_{\rm{f}}$ & $0.08\pm0.03$ & $0.08\pm0.04$\\ \\
	
Gaussian & E$_{line} $ (keV) &$6.3^f$ &$6.3\pm0.1$ \\
         & $\sigma $ (keV)&$0.4^f$ & $0.43^{+0.2}_{-0.1}$\\
         & norm ($\times 10^{-4}$) & $3.5\pm1.0$ & $5.0^{+2.0}_{-1.4}$  \\ \\
			
 NthComp1 & $\Gamma_1$ & $1.62\pm0.02$ & $1.63\pm0.02$ \\
	            & $kT_{\rm{e1}} $(keV) & $34.2^{+12.4}_{-6.1}$ & $ 39^{+23}_{-8}$\\
	            & norm & $0.14\pm0.01$ & $0.14\pm0.01$ \\ \\
             
  NthComp2 & $\Gamma_2$ & $2.0\pm0.1$ & $2.1^{+0.1}_{-0.2}$\\
	            & $kT_{\rm{e2}}$(keV) & $2^f$ & $2.05^{+0.78}_{-0.50}$ \\
                & norm ($10^{-2}$) & $4.2\pm0.9$ & $4.0^{+1.9}_{-1.6}$ \\ \\
                    \hline
    Cross-calibration & $C_{\rm{XRT/NICER}}$ & $0.97\pm0.02$ & $0.93\pm0.01$  \\ 
                          & $C_{\rm{NuSTAR/FPMB}}$ & $1.010\pm0.004$ & $1.010\pm0.004$ \\
                          \hline
                $\chi^2/dof$ & & $563.2/481$ & $548.5/582$ \\
              \hline
              & & \hspace{-3.8cm}Derived Values & \\
              \hline
   ${\rm{Flux_{unabs}}} [10^{-9} \mathrm{erg~cm}^{-2} \mathrm{s}^{-1}]$  & & $3.1$ & $3.3$ \\                
 ${\rm{Flux_{disk}}} [10^{-11} \mathrm{erg~cm}^{-2} \mathrm{s}^{-1}]$  & & $3$ & $16.3$ \\  
  ${\rm{Flux_{Soft}}}  [10^{-10} \mathrm{erg~cm}^{-2} \mathrm{s}^{-1}]$  & & $1.9$ & $1.7$ \\ 
  ${\rm{Flux_{Hard}}} [10^{-9} \mathrm{erg~cm}^{-2} \mathrm{s}^{-1}]$  & & $2.8$ & $2.8$\\ 
                Disk fraction ($\%$) & & $\sim1$ & $\sim5$\\
                Soft Comptonisation fraction ($\%$) & &$\sim6.1$ & $\sim5.2$\\
                 Hard Comptonisation fraction ($\%$) & & $\sim90.3$ & $\sim84.8$\\
                 & $\tau_1$ & $\approx3.2$ & $\approx2.9$ \\
                 & $\tau_2$ & $\approx12.4$ & $\approx11.5$ \\
                 \hline
    \end{tabular}
    \label{tab:Model-1}
    \begin{tablenotes}
    \item Note -- Flux is calculated in the $0.5-78$ keV band.
    \end{tablenotes}
\end{table*}


\section{Analysis and Results}
\subsection{Spectral Analysis} \label{sec:spectral}
We carried out broadband spectral analysis of the quasi-simultaneous Swift/XRT ($0.5-10$ keV), NICER ($0.5-10$ keV), and NuSTAR ($4-78$ keV) spectral data using XSPEC \citep[V.12.14.0;][]{Arnaud1996}. We quote the errors in the best-fit spectral parameters at the $90\%$ ($\Delta \chi^2 \approx2.71$) confidence level unless otherwise specified. For Obs1, we performed joint spectral modeling using Swift/XRT and NuSTAR data, while we used NICER and NuSTAR data for Obs2. We binned the spectral data from each instrument with a minimum of 25 counts per grouped bin using the optimal binning algorithm \citep{kaastra2016} in order to apply the $\chi^{2}$ statistics. We multiplied a \texttt{constant} factor to our models to account for differences in relative normalizations between the different instruments.

\begin{figure*}
\plottwo{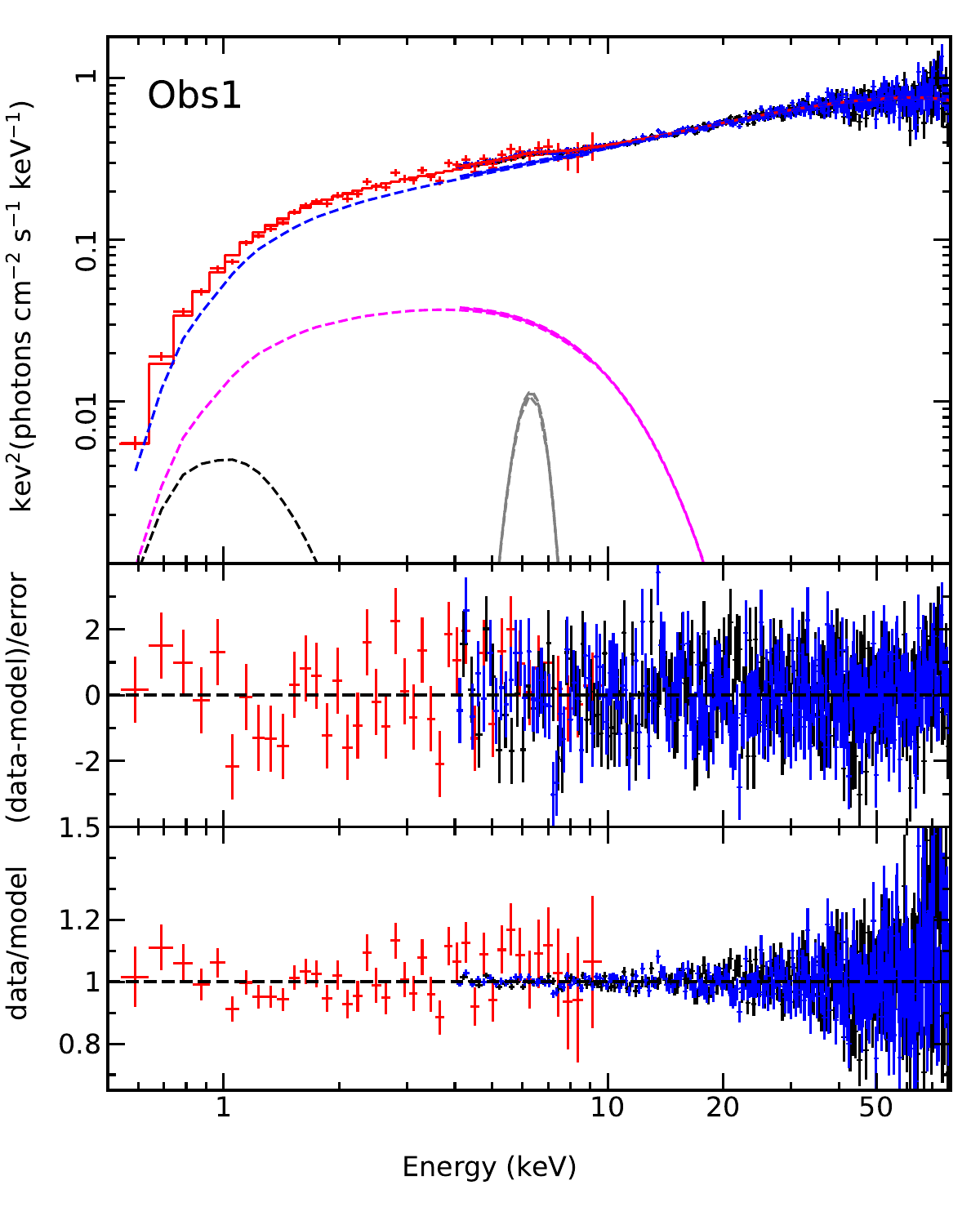}{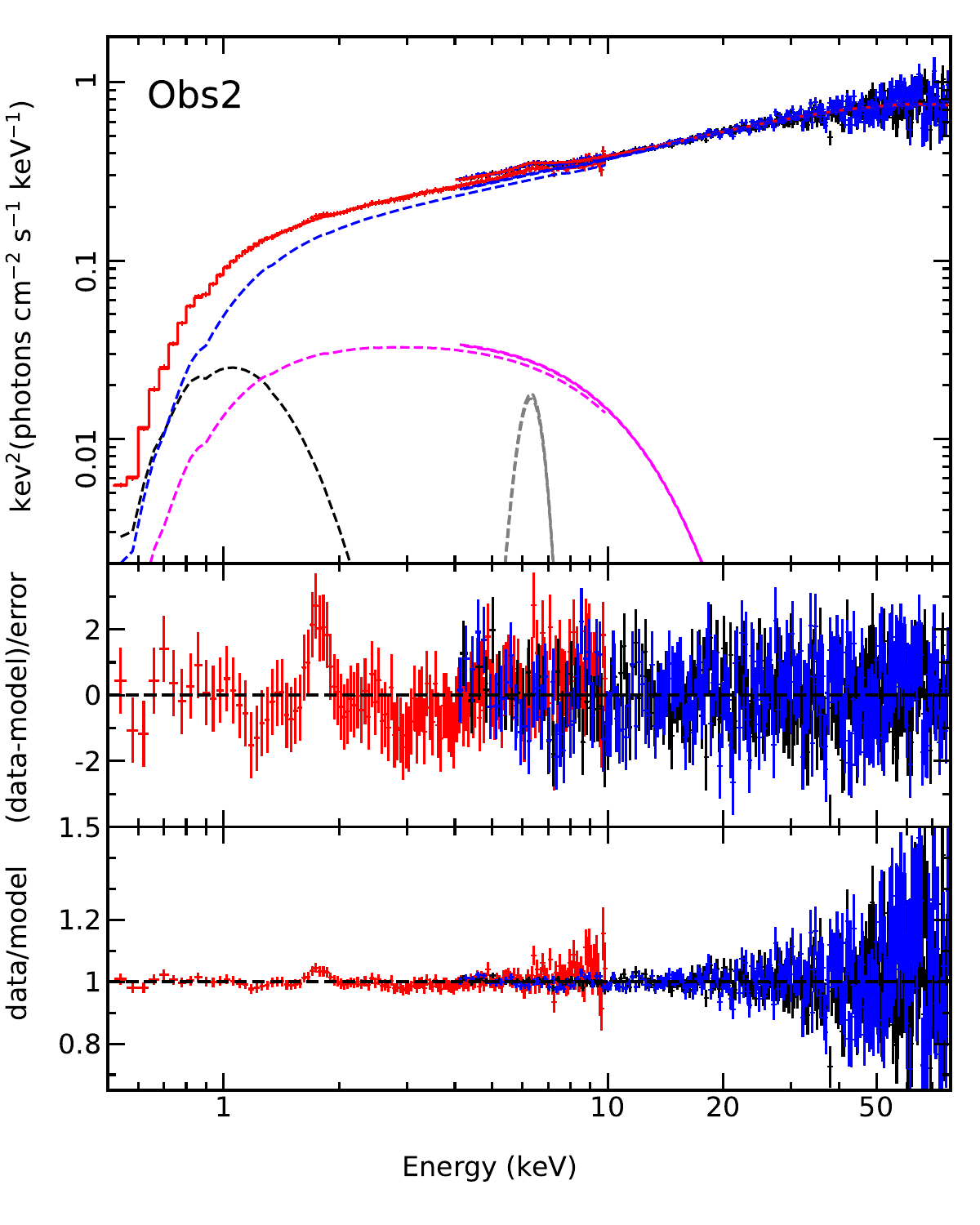}
\caption{Left panel: Joint Swift/XRT (red), Nustar/FPMA (black), and FPMB (blue) spectral data fitted with Model-1. Right panel: Joint NICER (red), NuSTAR/FPMA (black), and FPMB (blue) spectral data fitted with the same model. In both panels, solid lines represent the extrapolated models of the continuum fit. The dashed blue and magenta curves denote the hard and soft Comptonizing components, respectively. The black dashed curve corresponds to the disk emission, while the grey dashed line indicates the Gaussian component.}
 \label{fig:mod1} 
\end{figure*}

Initially, for both observations, we used a \texttt{powerlaw} model modified by the Galactic absorption component \texttt{tbfeo}, which also allows both oxygen and iron abundances to vary. The absorption cross-sections were taken from \citet{Verner1996ApJ...465..487V}, and the elemental abundances were adopted from \citet{Wilms2000ApJ...542..914W}. Given the higher spectral quality of the NICER data compared to that of the short-exposure Swift/XRT data, we first modeled Obs2 to constrain the oxygen abundance in the \texttt{tbfeo} model. Due to difficulty in constraining the iron abundance, we allowed it to vary over the range $0.5-2$. These values of the oxygen and iron abundances were then held fixed during the spectral modeling of Obs1. We also added a multi-color disk blackbody component \cite[\texttt{diskbb};][]{Mitsuda1984PASJ...36..741M} to our model to account for the thermal emission from the accretion disk. For Obs1, we fixed the inner disk temperature at $0.2$ keV due to the poor data quality, and kept the normalization parameter free to vary. However, this model was unable to describe the spectral data and provided poor fit for both the observations, leaving prominent residuals around the iron $K_\alpha$ ($\sim6.4$ keV) line and reflection hump ($\sim20-30$ keV). Figure~\ref{fig:fpma_res} shows the deviations of the NuSTAR/FPMA spectral data for both the observations from the best-fit model \texttt{tbfeo*(diskbb+powerlaw)}. We noticed prominent excesses due to iron emission line and reflection hump. We then replaced the \texttt{powerlaw} component with the \texttt{nthcomp} model, which accounts for the thermal Comptonization of the soft seed photons from the accretion disk in the hot corona. The \texttt{nthcomp} model is parameterized by the photon index ($\Gamma$) and electron temperature ($kT_{\rm{e}}$). We tied the seed photon temperature of the \texttt{nthcomp} component with the inner disk temperature ($kT_\mathrm{in}$) of the \texttt{diskbb} component. The model \texttt{constant*tbfeo*(diskbb+nthcomp)} provided an improved, though still unsatisfactory, fit for both the observations, with $\chi^{2}/$dof $= 933.6/484$ and $978.4/589$ for Obs1 and Obs2, respectively. We then incorporated the \texttt{ireflect} \citep{1995MNRAS.273..837M} component and convolved it with \texttt{nthcomp} to account for the reflection from ionized material. Considering the fact that \texttt{ireflect} is a convolution model, we extended the sample energy range from $0.01$ to $1000$ keV by using the \texttt{energies} command in \texttt{XSPEC}. We allowed the reflection scaling ($R_{\rm{f}}$) factor to vary while keeping all other parameters fixed, including the abundances of iron and other elements at their solar values, the inclination angle at $30^\circ$ \citep{Yao_2021_ApJ_920_121}, and the disk temperature at $10^6$ K. The disk ionization parameter was pegging to a very small value, and therefore we fixed it at $50$. Since \texttt{ireflect} does not consider the emission lines, we included a \texttt{gaussian} component to account for the iron $K_\alpha$ line. As mentioned earlier, we allowed the centroid line energy and width of the iron line to vary freely for Obs2, and then fixed the best-fit values in Obs1. This model \texttt{constant*tbfeo(diskbb+ireflect*nthcomp+gaussian)} improves the fit with $\chi^{2}/$dof $= 666/483$ and $623/585$ for Obs1 and Obs2, respectively.

However, the above model could not constrain the electron temperature ($kT_e$), which was pegged to the highest model-defined value of $1000$ keV. Besides, significant residuals persisted at $\leqslant10$ keV, suggesting that the coronal structure may not be homogeneous and an additional Comptonizing region is required to fit the broadband spectral data adequately. It is worth mentioning here that \citet{Yao_2021_ApJ_920_121} studied AT2019wey during its initial discovery outburst in 2019, when the source was in the LHS. The authors employed a single Comptonizing region in their spectral model but were unable to constrain the coronal electron temperature, which they fixed at $1000$ keV. Apart from this, \cite{Feng2022SCPMA..6519512F} also performed X-ray spectral analysis using NuSTAR observations by keeping the electron temperature or cut-off energy in their respective model components fixed at $100$ keV or $300$ keV, respectively. This further suggests that the structure of the corona may be inhomogeneous, and modeling the corona as a single Comptonizing medium results in the electron temperature pegging at the highest model-defined values. Hence, we incorporated another \texttt{nthcomp} component, independent of the first one, into our model. Similar to the first \texttt{nthcomp}, we considered the seed photon temperature of the second Comptonizing region the same as the disk photons. Other parameters, such as photon index, electron temperature, and normalization of this component, were allowed to vary independently from the first Comptonizing component. Addition of this second \texttt{nthcomp} component provided a substantial improvement in the fit statistics and our model \texttt{constant*tbfeo(diskbb+ireflect*nthcomp+gaussian+\hspace{0.5 cm}nthcomp)} (hereafter Model-1) now offers a satisfactory and statistically acceptable fit to the broadband spectral data with $\chi^{2}/$dof$ = 563.2/481$ and $548.5/582$ for Obs1 and Obs2, respectively. The best-fit spectral parameters obtained from Model-1 are listed in Table~\ref{tab:Model-1}, whereas the best-fit spectral models along with data are shown in Figure~\ref {fig:mod1}.

We find the best-fit Galactic absorption column density ($N_{\rm{H}}$) to be $\sim0.6\times10^{22}~\rm{cm^{-2}}$ for the both observations, which is consistent with the value estimated by \cite{Yao_2021_ApJ_920_121}. The inner disk temperature obtained from Obs2 is $\sim0.2$ keV, which closely matches the assumed value for Obs1. For the first Comptonizing region, the best-fit photon index ($\Gamma$) and electron temperature ($kT_{\rm{e}}$) are $\sim1.6$ and $\sim39$ keV, respectively, similar to the values typically observed in the LHS of BHXRBs. On the other hand, the second Comptonizing region appears to be significantly cooler, with $kT_{\rm{e}} \sim2$ keV and a steeper spectrum ($\Gamma \sim2$) compared to the first region. The reflection scaling ($R_{\rm{f}}$) factor from the ionized material is found to be consistent within the uncertainties across the two observations. The centroid energy and full width at half maximum (FWHM) or $\sigma$ of the iron K$_\alpha$ line are measured to be $\sim6.3$ keV and $\sim0.4$ keV, respectively. Furthermore, we find that the hard Comptonizing region dominates the overall flux, contributing $\gtrsim 85\%$ in both observations, while the soft Comptonizing region and the disk contribute only a few percent. This is also prominent from the normalization of the \texttt{nthcomp} component associated with the hard Comptonizing region, which is an order of magnitude higher than that of the soft Comptonizing region. We also derived the Thomson optical depths ($\tau$) for the two Comptonizing regions from the best-fit values of $\Gamma$ and $kT_{\rm{e}}$ using the following relation \citep{Zdziarski1996MNRAS.283..193Z, Zycki1999MNRAS.309..561Z}:
\begin{equation}
    \tau = \sqrt{\frac{9}{4} + \frac{3m_e c^2}{kT_e [(\Gamma + \frac{1}{2})^2 - \frac{9}{4}]}} -\frac{3}{2}
\end{equation}
where $c$ is the speed of light and $m_e$ is the mass of the electron. The derived values of $\tau$ are $\sim3$ and $\sim12$ for the hard and soft Comptonizing regions, respectively, implying that the soft Comptonizing region is highly optically thick in comparison to the hard region.

\begin{table*}
  \caption{Best-fit broadband X-ray spectral parameters derived using Model-2}
   \renewcommand{\arraystretch}{1.0}
    \centering
  \begin{tabular}{lcccc}
	\hline
    \hline
    Components & Parameter & Obs1 & Obs2\\
   \hline  
    TBfeo & $N_H$ $(\times 10^{22}$ cm$^{-2}$) & $0.66^{+0.07}_{-0.13}$ & $0.64^{+0.02}_{-0.01}$ \\
            & $O$ & $1.28^f$ &  $1.28\pm0.08$  \\
            & $Fe$  & $0.5^f$ & $0.50^{+0.09}_{}$ \\ \\

    Diskbb & $kT_{\mathrm{in}}$(keV) & $0.2^f$ & $0.22\pm0.01$  \\
            & norm & $19139^{+6760}_{-5567}$ & $16607^{+5488}_{-4717}$ \\ \\
   
    Thcomp & $\Gamma_1$ & $1.64\pm0.01$  & $1.62^{+0.03}_{-0.04}$ \\
           & $kT_{\rm{e1}}$ (keV)& $43.6^{+36.0}_{-10.5}$ & $52.3^{+317.8}_{-14.9}$ \\
           & $C_{\rm{f}}$ & $0.5^{+0.2}_{-0.1}$ & $0.34^{+0.15}_{-0.10}$  \\ \\

    RelxillCp$_1$ & $q$ & $3^f$ & $3^f$ \\
                & $a$  & $0.998^f$ & $0.998^f$ \\
                & $i^\circ$ & $30^f$ & $34.6^{+6.1}_{-5.2}$  \\
                & A$_{\mathrm{Fe}}$ & $1^f$ & $1^f$  \\
                & $R_{\rm{in}}~(r_{\mathrm{g}})$ & $27.5^{+28.9}_{-11.1}$ & $25.6^{+14.2}_{-7.9}$\\
                & $\log N$ & $19^f$ & $19.0^{+0.3}_{-0.6}$\\
                & $\log \xi$ & $2.5^{+0.3}_{-2.0}$ & $1.7^{+0.3}_{-0.2}$ \\
                & $R_{\rm{f1}}$ & $-1^f$ & $-1^f$  \\
                & ${\rm{norm_1}}$ ($\times10^{-4}$)  & $4.5^{+2.7}_{-1.8}$ & $9.4^{+4.6}_{-3.3}$ \\ \\

    mbknpo      & $B$ & $0.47^f$  & $0.59^f$ \\
                & $I$ & $0.84^f$ & $0.82^f$\\ \\
    
    RelxillCp$_2$  & $\Gamma_2$ & $1.3^{+0.3}$ & $1.79^{+0.11}_{-0.15}$ \\
                & $kT_{e2}$ (keV)  & $1.3^{+0.2}_{-0.1}$ & $2.1^{+0.6}_{-0.3}$  \\
                & $R_{\rm{f2}}$ & $1^f$ & $0.9^{+1.0}_{-0.6}$ \\
                & ${\rm{norm_2}}$ ($\times10^{-4}$) & $0.7^{+1.3}_{-0.2}$ & $4.3^{+2.5}_{-2.2}$ \\ \\

    mbknpo      & $B$ & $0.47^f$ & $0.59^f$\\
                & $I$ & $0.5^f$ & $0.99^f$\\ 
                \hline
                 Cross-calibration & $C_{\rm{XRT/NICER}}$ & $1.0^{+0.02}_{-0.03}$ & $0.939\pm0.006$  \\ 
                          & $C_{\rm{NuSTAR/FPMB}}$ & $1.014\pm0.004$ & $1.01\pm0.004$ \\
                \hline
                $\chi^2/dof$ & & $532.4/479$ & $508.7/580$  \\
                \hline
              & & \hspace{-3.8cm}Derived Values & \\
              \hline
   ${\rm{Flux_{unabs}}} [10^{-9} \mathrm{erg~cm}^{-2} \mathrm{s}^{-1}]$  & & $3.3$ & $3.4$ \\                
 ${\rm{Flux_{disk}}} [10^{-10} \mathrm{erg~cm}^{-2} \mathrm{s}^{-1}]$  & & $2.3$ & $3.1$ \\  
  ${\rm{Flux_{Soft}}}  [10^{-10} \mathrm{erg~cm}^{-2} \mathrm{s}^{-1}]$  && $1.1$ & $3.6$ \\ 
  ${\rm{Flux_{Hard}}} [10^{-9} \mathrm{erg~cm}^{-2} \mathrm{s}^{-1}]$  & & $3.0$ & $2.7$\\ 
                Disk fraction ($\%$) & & $\sim6.9$ & $\sim9.1$\\
                Soft Comptonisation fraction ($\%$) & &$\sim3.2$ & $\sim10.7$\\
                 Hard Comptonisation fraction ($\%$) & & $\sim90$ & $\sim80.3$\\
  \hline
\end{tabular}
\label{tab:Model-2}
\begin{tablenotes}
    \item Note -- Flux is calculated in the $0.5-78$ keV band.
    \end{tablenotes}
\end{table*}

We also then tested a more accurate model \texttt{relxillCp} and carried out reflection spectroscopy. The \texttt{relxillCp} model belongs to the \texttt{relxill} family of relativistically blurred reflection models \citep{Dauser2014MNRAS.444L.100D, Garcia2014ApJ...782...76G}. Notably, the incident continuum in \texttt{relxillCp} is identical to \texttt{nthcomp}. We first tried to fit the broadband spectra with a single \texttt{relxillCp} by replacing the \texttt{nthcomp}, \texttt{ireflect}, and \texttt{gaussian} components from our earlier model. In this setup, we used the \texttt{relxillCp} as a reflection component only and incorporated the convolution model \texttt{thcomp} \citep{2020MNRAS.492.5234Z}, which is an improved version of \texttt{nthcomp}, to describe the incident radiation. We convolved \texttt{thcomp} with \texttt{diskbb} component and tied the common parameters between \texttt{thcomp} and \texttt{relxillCp}. We fixed the black hole spin parameter in \texttt{relxillCp} at the maximum value \citep{Feng2022SCPMA..6519512F}, and adopted a single emissivity profile ($\propto r^{-q}$, $q$ being the emissivity index) across the entire accretion disk by tying the break radius to the outer disk radius. Allowing $q$ to vary resulted in the parameter pegging at the lowest defined bound of 3. We, therefore, fixed $q$ at $3$ during the spectral fitting. We also fixed the iron abundance at the solar value for both the observations. 

\begin{figure*}
\plottwo{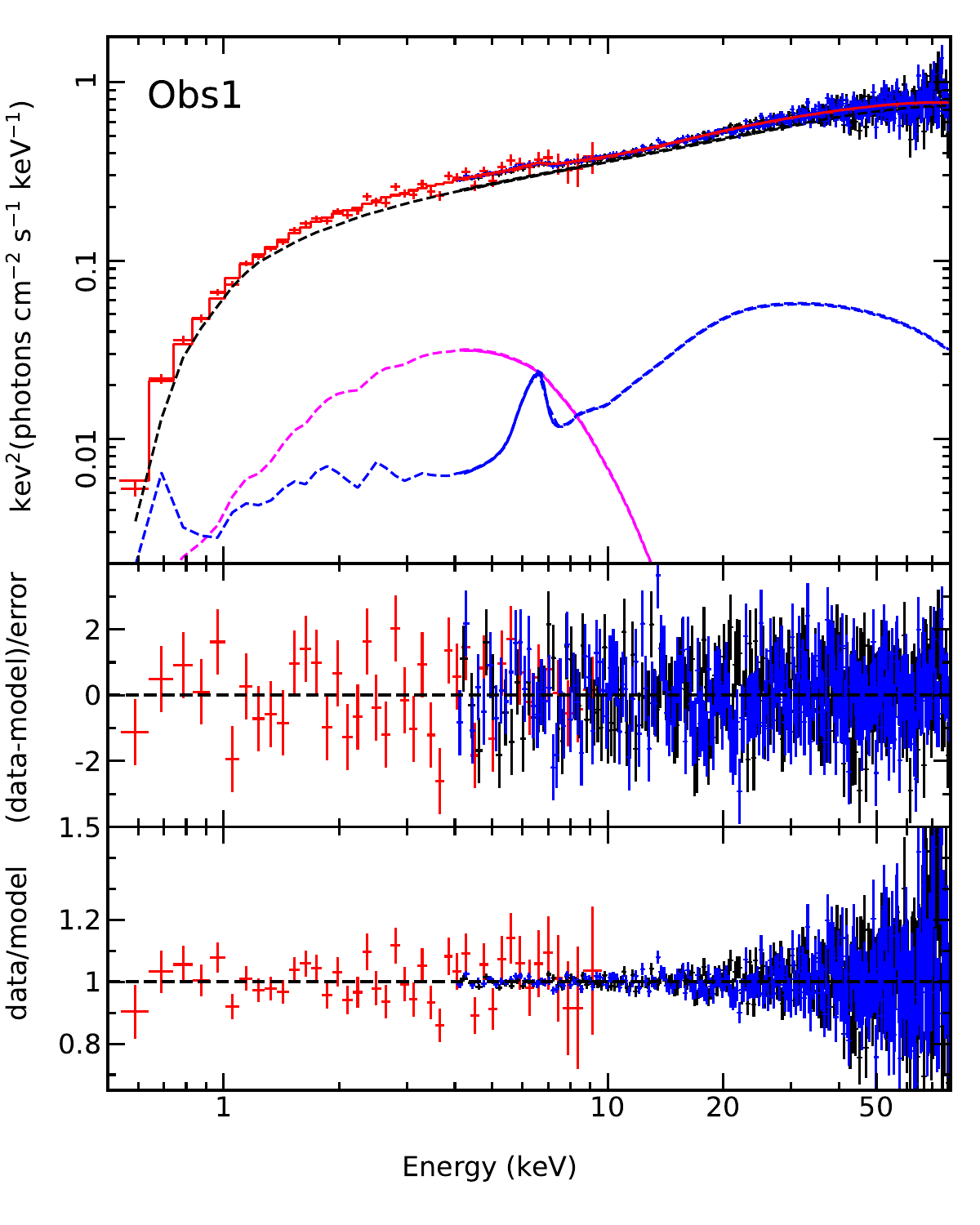}{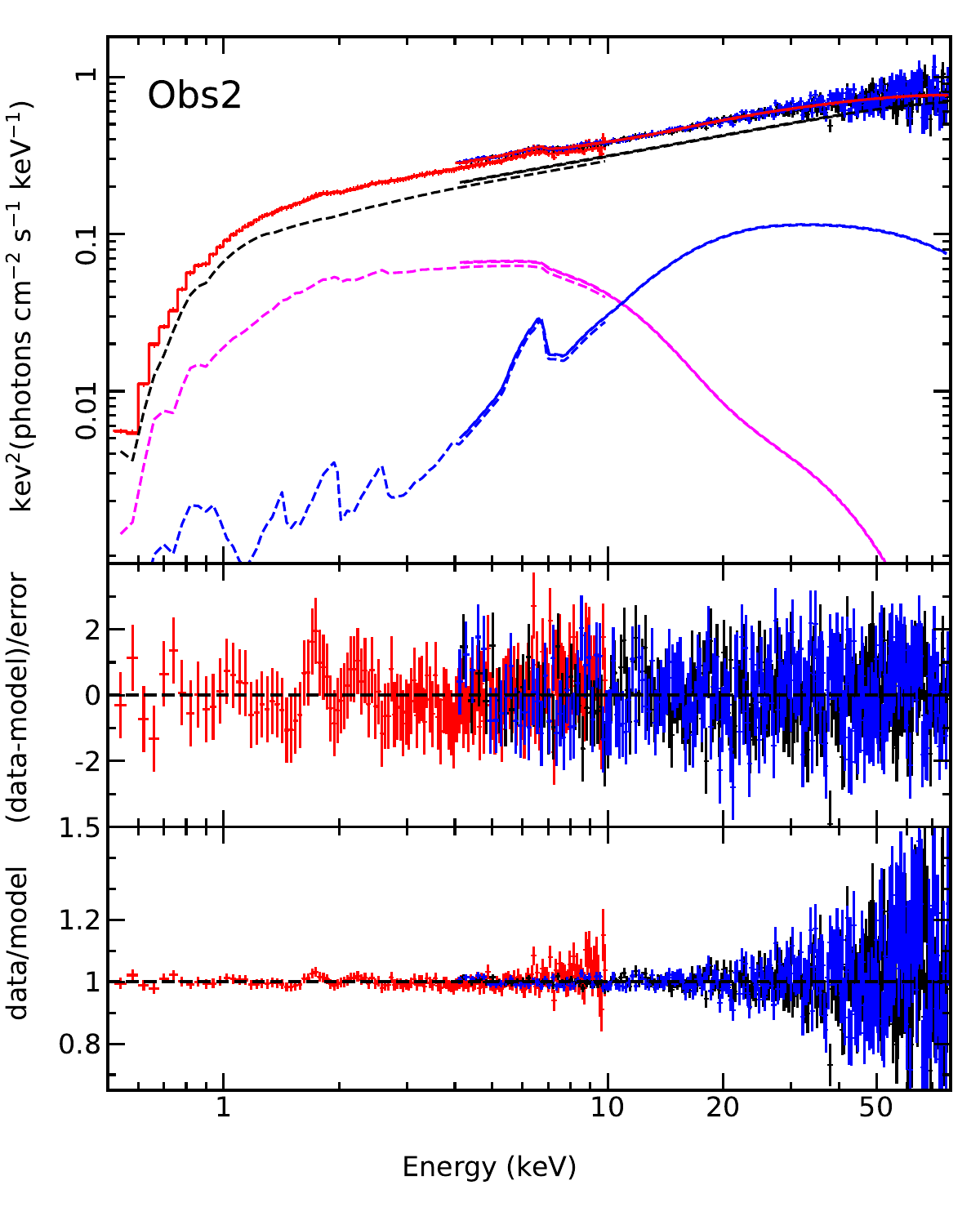}
\caption{Left panel: Joint Swift/XRT (red), Nustar/FPMA (black), and FPMB (blue) spectral data fitted with Model-2. Right panel: Joint NICER (red), NuSTAR/FPMA (black), and FPMB (blue) spectral data fitted with the same model. In both panels, solid lines represent the extrapolated models of the continuum fit. The black dashed curve shows the thermal Comptonization from the hard Comptonizing region, while the blue dashed curves represent the corresponding reflection component. The magenta dashed curve denotes the thermal Comptonization and associated reflection from the soft Comptonizing region.}
\label{fig:Mod-2}
\end{figure*}

Other free parameters in \texttt{relxillCp} are inner disk radius ($\rm{R_{in}}$), log of disk ionization ($\log \xi$) and density ($\log N$), inclination angle ($i$) and normalization. However, we fixed $i$ at $30^\circ$ in Obs1 due to the poor quality of the data. Besides, to mitigate the overprediction of the reflection component at low energies, we multiplied the \texttt{relxillCp} model with the \texttt{mbknpo} component. This multiplicative model \texttt{mbknpo} introduces a break in the reflection spectrum below a specified energy, thereby suppressing the unphysical rise of the photon flux at low energies \citep{Steiner2024ApJ...969L..30S, Svoboda_2024}. This correction is governed by two parameters: the break energy $B$ and the spectral index $I$, which modify the reflection spectrum for energies $E<B$. We fixed the parameter $B$ at its best-fit value, and set $I$ to a fixed value given by $I=\Gamma - 0.8$. This model  \texttt{constant*tbfeo(thcomp*diskbb+mbknpo*relxillcp)} provided poor spectral fits with $\chi^2$/dof$=729.4/482$ and $679.8/584$ for Obs1 and Obs2, respectively. Similar to the case with a single \texttt{nthcomp}, the electron temperature, $kT_{\rm{e}}$, could not be constrained, and thus we incorporated another \texttt{relxillCp} component into our model. We considered the second \texttt{relxillCp} component to be responsible for both the irradiation and the associated reflection. We also multiplied another \texttt{mbknpo} model with the second \texttt{relxillCp} component. The common parameters, except the photon index ($\Gamma$), electron temperature ($kT_{\rm{e}}$), reflection fraction ($R_{\rm{f}}$), and normalization, were tied between the two \texttt{relxillCp} components. Since the hard Comptonizing region is found to be dominating the overall flux (see Table~\ref{tab:Model-1}), the inner disk radius ($R_{\rm{in}}$) was tied between the two regions, and hence the log of ionization ($\log \xi$). We also considered the disk densities corresponding to the reflections associated with both the Comptonizing regions to be the same. Inclusion of the additional \texttt{relxillCp} component provided a significant improvement in the fit-statistics for both the observations, and this model \texttt{constant*tbfeo(thcomp*diskbb+mbknpo*relxillcp+\hspace{0.5cm}mbknpo*relxillcp)} (referred as Model-2) provided very good fit with $\chi^{2}/$dof$=532.4/479$, and $508.7/580$, for Obs1 and Obs2, respectively. Notably, similar to the above, the abundances of the interstellar medium for Obs1 were fixed at the best-fit values obtained from Obs2, given the poor quality of data in Obs1. The best-fit spectral parameters are given in Table~\ref{tab:Model-2}, and the best-fit models along with the spectral data are shown in Figure~\ref{fig:Mod-2}. 

Our spectral results from Model-2 suggest that the characteristics of the two distinct Comptonizing regions appear to be very similar to what we obtained in our Model-1. The inner disk temperature of $\sim0.2$ keV is consistent between both the observations and models as well. We find the inner accretion disk to be truncated away from the ISCO by tens of $r_{\rm{g}}$ for both the observations. The inclination angle, $i$, obtained from the Obs2 is either consistent or very close to the previously reported values.  A high disk density of $10^{19}~\rm{cm^{-3}}$ is measured in Obs2 and was subsequently fixed during the spectral modeling of Obs1. The ionization parameter of the disk is found to be low from Obs2, however, it is poorly constrained in Obs1 due to the poor quality of the Swift/XRT data. As in Model-1, we find that the hard Comptonizing region dominates the emission, contributing $\sim 80-90 \%$ of the observed flux.

\subsection{Timing Analysis}

\subsubsection{PDSs}

We derived rms-normalized and Poisson noise subtracted PDSs in the $0.5-10$ keV band using Swift/XRT and NICER observations, and cross-power density spectra (CPDSs) in the $10-30$ keV band using NuSTAR observations. For the Swift/XRT and NICER observations, we first derived lightcurves in the $0.5-10$ keV band with a time bin size of $0.05$ s, corresponding to the Nyquist frequency of $10$ Hz. These lightcurves were then utilized to derive the PDSs using the \texttt{powspec} task, available in HEASoft as a part of the \texttt{XRONOS} package. We also derived the CPDSs from the NuSTAR observations in the $10-30$ keV band with the default time bin size embedded in the \texttt{HENDRICS} package \citep{Bachetti2015ApJ...800..109B}. We analyzed all the PDSs and CPDSs for the frequency range of $0.01-10$ Hz within \texttt{XSPEC}, and quoted the uncertainties in the $90\%$ confidence levels. Neither the PDSs nor the CPDSs exhibit any QPO like narrow and sharp features. Rather, they can be described well with a combination of two broad Lorentzians required for the band-limited noise (BLNs) components, typically observed in the PDSs of BHXRBs in the LHS \citep{Belloni2002ApJ...572..392B, Wilkinson2009MNRAS.397..666W}. 

In the case of Obs1, the PDS derived from the Swift/XRT observation appears to be noisy due to large uncertainties at high frequencies. A single Lorentzian component can sufficiently describe this PDS. On the other hand, a single Lorentzian could not describe the CPDS derived from the NuSTAR data in Obs1, and the presence of slight excesses was noticed in the data residuals. Addition of another Lorentzian improved the fit significantly with $\Delta \chi^2=-29.7$. However, we did not notice any such improvement for the PDSs from the Swift/XRT data. For Obs2, the PDS and the CPDS from the respective NICER and NuSTAR observations were best described with two broad Lorentzian components. All the best-fit temporal parameters are listed in Table~\ref{tab:pds_para}, and the best-fit PDSs and CPDSs are shown in Figure~\ref{fig:best-fit PDSs}.

\begin{figure*}
    \includegraphics[width=\columnwidth]{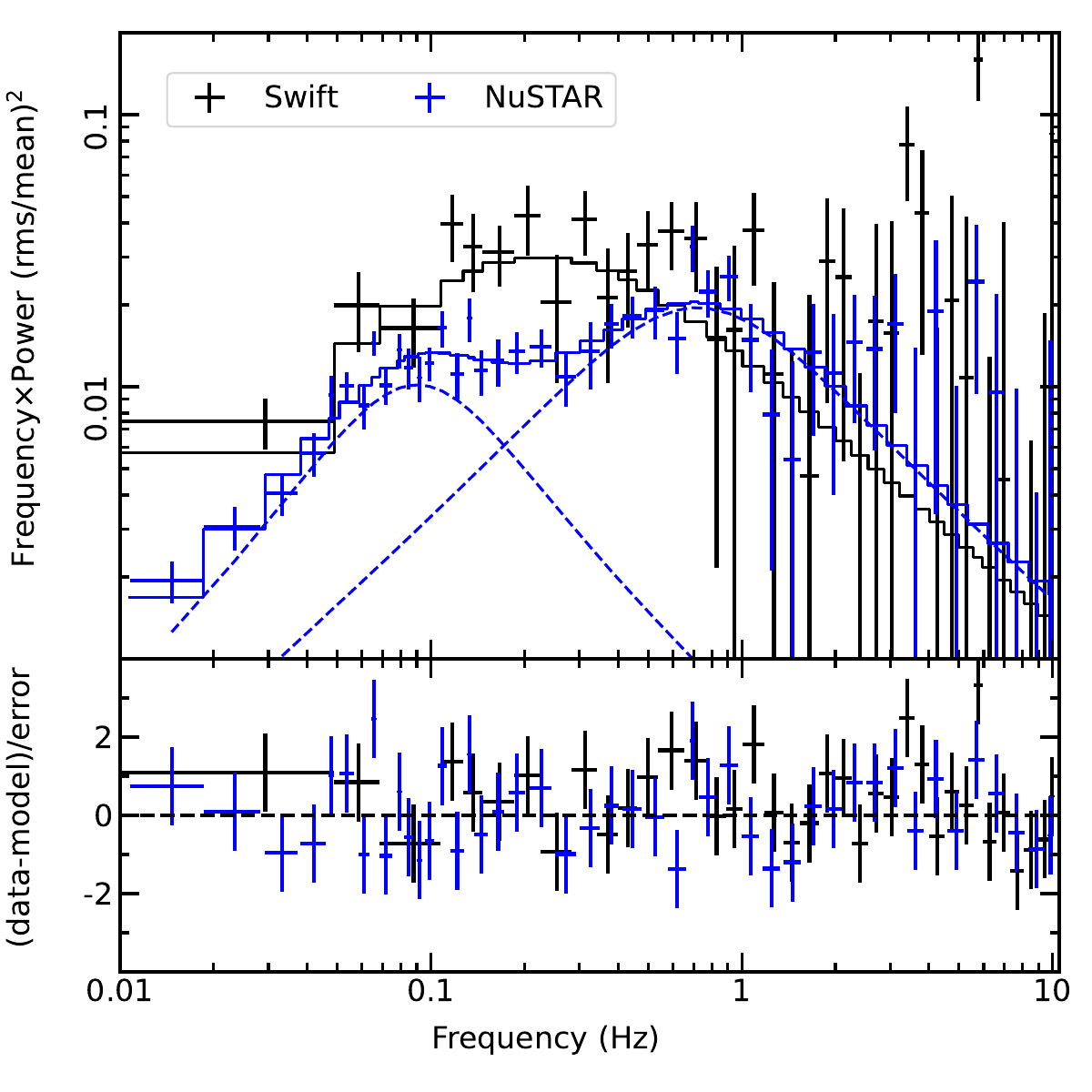}
    \includegraphics[width=\columnwidth]{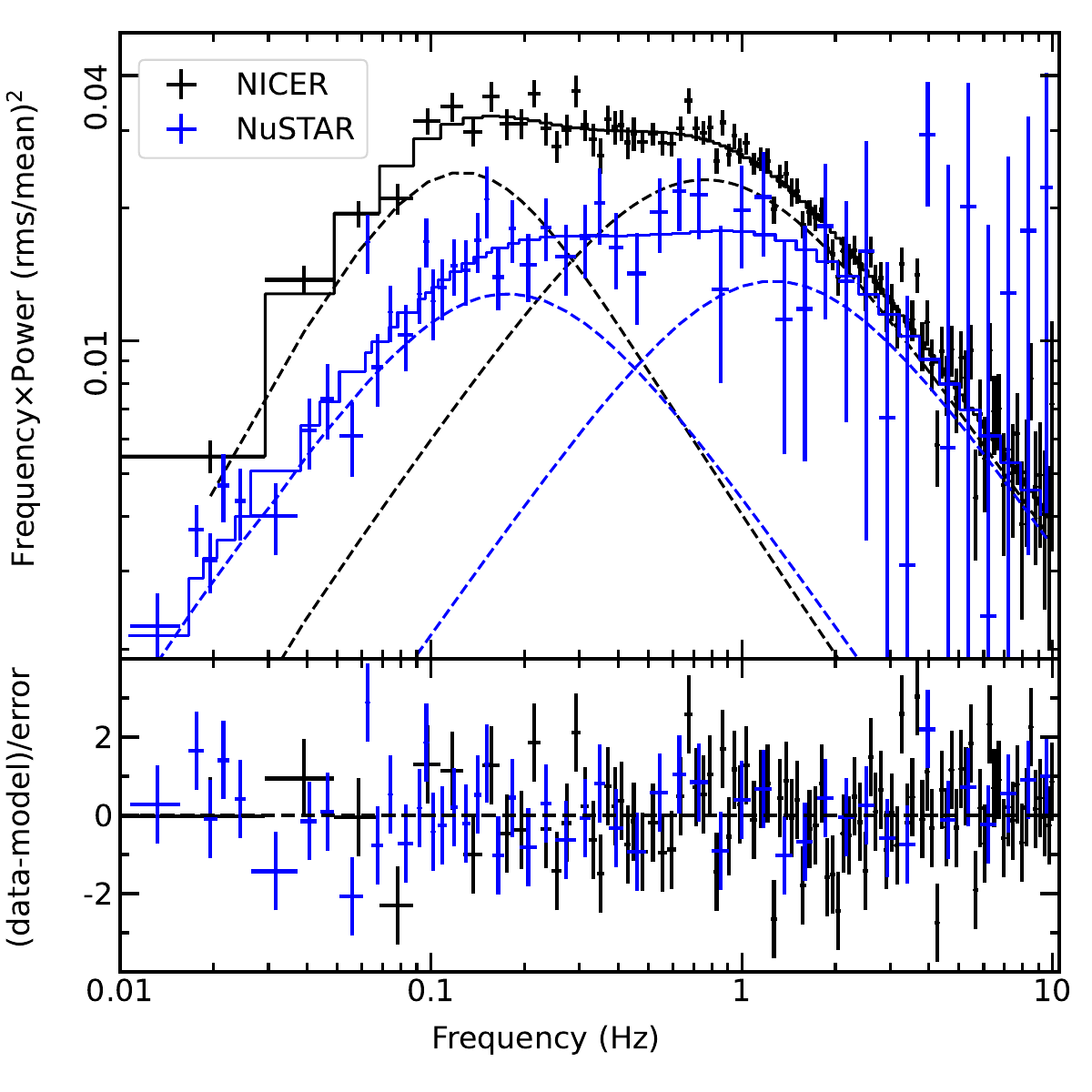}
   \caption{Left panel: PDS from Swift/XRT ($0.5-10$ keV) and CPDS from NuSTAR ($10-30$ keV) data. Right panel: PDS from NICER ($0.5-10$ keV) and CPDS from NuSTAR ($10-30$ keV) data. The PDSs are rebinned for plotting purposes only.}
    
    \label{fig:best-fit PDSs}
\end{figure*}

The shapes of the PDSs and CPDSs in different energy bands reveal that the variability power in the $0.5–10$ keV band remains comparatively higher than that in the $10-30$ keV band, particularly in the low-frequency regime. As the frequency increases, the power in both the bands becomes comparable. However, large uncertainties at high frequencies make it difficult to make a concrete statement on the variability power at these frequencies. We further examined the shapes of the PDSs in two distinct energy bands of $0.5–1$ keV, typically dominated by disk emission, and $5–10$ keV, dominated by coronal emission, using the NICER data in Obs2. A similar trend was observed, where the PDS in the $0.5–1$ keV band exhibited larger power at low frequencies compared to the PDS in the $5–10$ keV band. At high frequencies, the power spectra in both the bands gradually become similar within the uncertainty levels. 

 \begin{table*}[!ht]
\caption{Best-fit temporal parameters of the PDSs derived from Swift/XRT, NICER and CPDSs derived from NuSTAR data}
   \centering
   \setlength{\tabcolsep}{15pt}
   \renewcommand{\arraystretch}{1.2}
  \begin{tabular}{lccccc }
      \hline
      \hline
      &   \multicolumn{2}{c}{Obs1} & & \multicolumn{2}{c}{Obs2} \\
      \cline{2-3} \cline{5-6} 
    &   \multicolumn{1}{c}{Swift/XRT} &  {NuSTAR} & & {NICER} & {NuSTAR}\\
       \cline{2-3} \cline{5-6}
         
  Parameter & {$0.5-10$ keV} &{$10-30$ keV} & & {$0.5-10$ keV} & {$10-30$ keV} \\
  \hline
   
     $\nu_{\rm{bln1}}$(Hz) & $0^f$ & $0^f$ & & $0^f$ & $0^f$ \\
     $\rm{width_{bln1}}$(Hz) & $0.45^{+0.15}_{-0.11}$ & $0.35^{+0.08}_{-0.07}$ & & $0.23^{+0.04}_{-0.03}$ & $0.13^{+0.06}_{-0.05}$ \\
     $\rm{rms_{bln1}}$[\%] & $30.7^{+2.2}_{-2.4}$ &  $20^{+2.5}_{-2.7}$  & & $25.2^{+1.6}_{-1.4}$ & $14.3^{+2.7}_{-3.0}$\\ \\
     
     $\nu_{\rm{bln2}}$(Hz) & \nodata & $0^f$ & & $0^f$ & $0^f$ \\
     $\rm{width_{bln2}}$(Hz) & \nodata & $2.53^{+1.51}_{-0.98}$ & & $1.52^{+0.15}_{-0.10}$ & $1.29^{+0.54}_{-0.47}$ \\
     $\rm{rms_{bln2}}$[\%] & \nodata & $20.7^{+2.5}_{-5.4}$ & & $27^{+1.1}_{-1.5}$ & $22.2^{+3.0}_{-3.6}$ \\
     \hline
     
     $\chi^2$/dof & $110/94$ & $120/161$ & & $127.7/91$ & $177.4/160$ \\
         \hline 
   \end{tabular}
   \label{tab:pds_para}
\end{table*}


\subsubsection{Fractional rms variability amplitude}

We derived the fractional rms variability amplitude for the $0.01-10$ Hz frequency range as a function of photon energy using all the observations by Swift/XRT, NICER, and NuSTAR. We used the \texttt{Stingray} \citep{Huppenkothen2019JOSS....4.1393H, Huppenkothen2019ApJ...881...39H} package to derive the rms-energy spectra from the Swift and NICER observations. For the NuSTAR observations, we first extracted the CPDSs at different energy bands and then calculated the fractional rms in the $0.01-10$ Hz frequency range for each CPDS. Figure~\ref{fig:rms_E} exhibits the rms-energy spectra for both the observations. We restricted our fractional rms variability amplitude analysis up to $5$ keV and $30$ keV for the Swift/XRT and NuSTAR data, respectively, to avoid the large uncertainties at high energies due to low S/N ratio. The rms-energy spectra from both the observations exhibit quite similar characteristics, with the variability being higher at lower energies and gradually decreasing toward higher energies. Notably, the slightly lower fractional rms observed in the NuSTAR data, compared to those from other instruments in the same energy bands, may be attributed to the detector deadtime effect, which can artificially suppress the measured variability amplitude \citep[M. Bachetti, private communication;][]{Bachetti2018ApJ...853L..21B, Zdziarski2024ApJ...967L...9Z}. This effect becomes increasingly significant at high source fluxes. However, given the relatively low flux of the source in our observations, we find that the impact of NuSTAR deadtime on the fractional rms is minimal. Consequently, the rms values obtained from NuSTAR data agree closely with those measured with Swift/XRT and NICER within the uncertainty levels.

\begin{figure*}
    \includegraphics[width=\columnwidth]{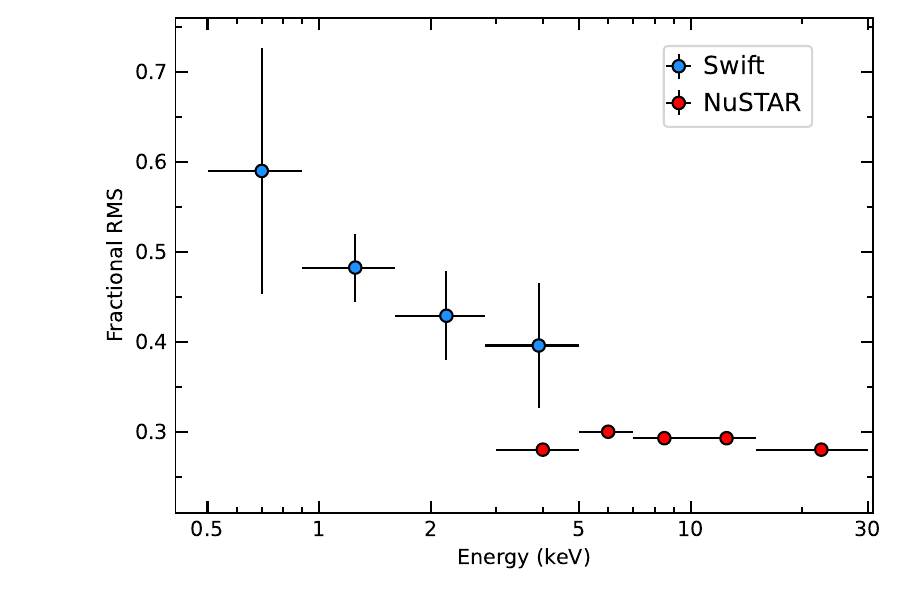}
    \includegraphics[width=\columnwidth]{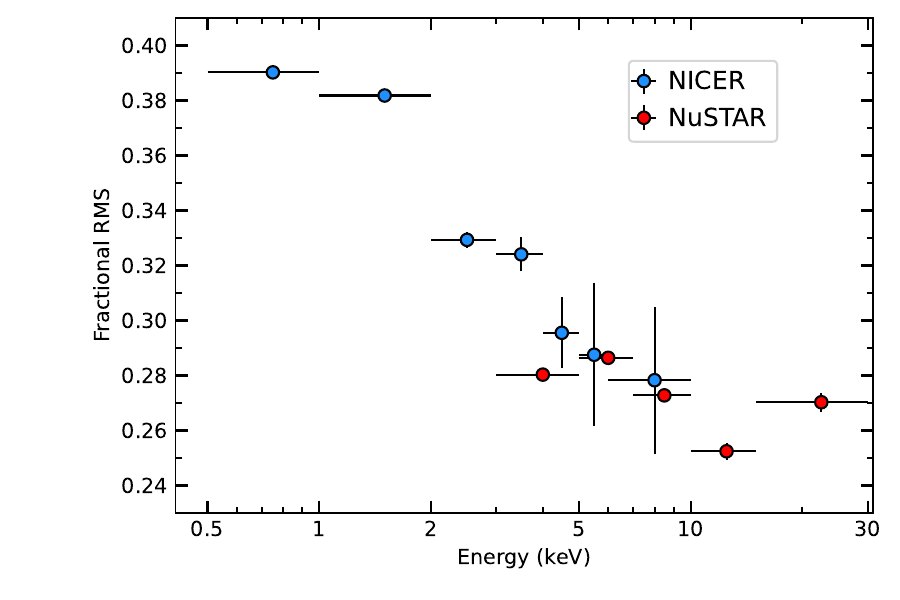}
    \caption{Fractional rms variability amplitude as a function of photon energy derived for the frequency range of $0.01-10$ Hz for Obs1 (left panel) and Obs2 (right panel).}
    \label{fig:rms_E}
\end{figure*}

\subsubsection{Time lags}
\paragraph{Lag-frequency spectrum}
Study of time lags provides crucial insights into the nature of variability in the source. We restricted our lag analysis to the NICER data from Obs2 only, as the Swift/XRT data have a very poor S/N ratio due to short exposure. We first derived time lags as a function of Fourier frequency between two different energy bands of $0.5-1.5$ keV and $1.5-10$ keV using the \texttt{Stingray}  \citep{Huppenkothen2019JOSS....4.1393H, Huppenkothen2019ApJ...881...39H} package, following the cross-spectral analysis methods prescribed by \citet{Vaughan1997ApJ...474L..43V} and \citet{Nowak1999ApJ...510..874N}. The time lags were calculated considering a time bin size of $0.05$ s, and the length of each segment as $100$ s, which enabled us to probe the nature of lag-frequency spectrum in the frequency range of $0.01-10$ Hz. We detect the presence of prominent hard or positive lags at the low frequency bins, indicating that photons in the high-energy band of $1.5-10$ keV are lagging behind the photons in the relatively lower energy band of $0.5-1.5$ keV. Previous studies have shown that the amplitude of hard lags decreases with increasing frequency, and that the lag-frequency spectrum in the $0.05-0.6$ Hz can be well described by a power-law model, with the slope ranging from $\sim -0.7$ to $\sim -1.1$ \citep{Nowak1999ApJ...510..874N, DeMarco2015ApJ...814...50D}. Here, we also fitted the lag-frequency spectrum with a power-law model in the $0.05-0.6$ Hz frequency range to understand its nature. The obtained best-fit power-law slope is found to be $-0.83\pm0.21$, which agrees closely with the range mentioned above. The upper panel of Figure~\ref{fig:lag_freq} shows the derived time lags as a function of frequency and the best-fit power-law model. As can be seen from this figure, the lag-frequency spectrum in the high-frequency bins closely follows the best-fit power-law model. However, a significant drop in the amplitude of the hard lags is observed towards the low frequency bins at $\sim0.2$ Hz and  $\lesssim0.1$ Hz.

\begin{figure}
    \includegraphics[width=\columnwidth]{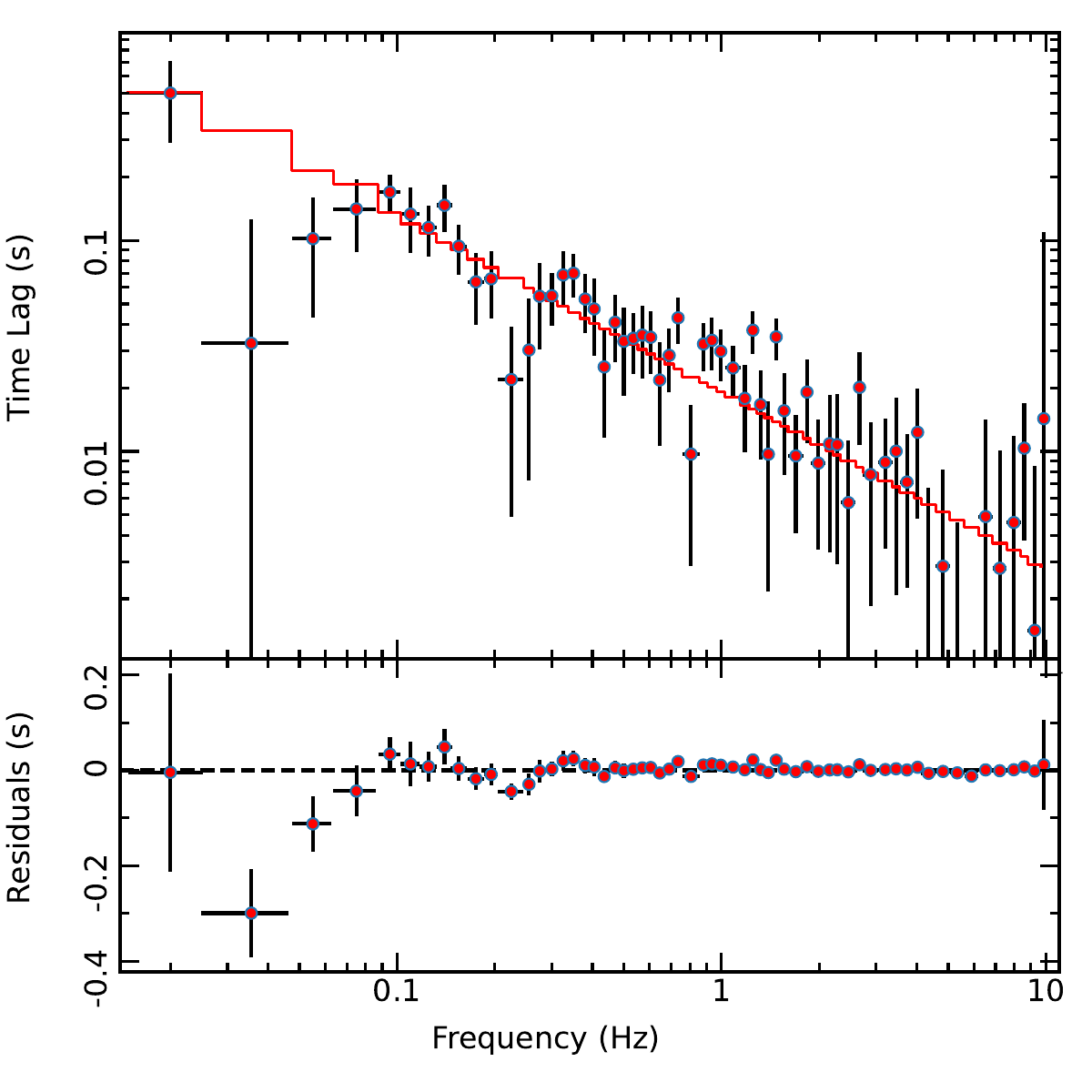}
    \caption{Upper panel: Time lags as a function of Fourier frequency computed between $0.5-1.5$ keV and $1.5-10$ keV bands, and the best-fit power-law model. Lower panel: Time lags residuals to the best-fit powerlaw model. The lag-frequency spectrum is rebinned for plotting purposes only.}
    \label{fig:lag_freq}
\end{figure}

\paragraph{Frequency resolved lag-energy spectrum}
We then derived frequency-resolved lag-energy spectra to probe the energy dependence of the observed time lags. For this, we computed time lags across a series of adjacent energy bands relative to a reference band. We choose the reference band as the full energy band ($0.5-10$ keV), excluding the energy bin of our interest. Using such a broad energy band as the reference band helps to achieve a high signal-to-noise ratio in the lag-energy spectrum, and reduce Poisson noise \citep{Uttley2014A&ARv..22...72U}. The time lags were averaged over the frequency range of $0.02-0.2$ Hz, where a significant drop in the hard lag amplitude was noticed (see Figure~\ref{fig:lag_freq}), and the coherence was also found to be high ($\gtrsim0.8$). Here, a positive lag for a given energy band indicates that photons in that band are delayed relative to the rest of the energy bands. Figure~\ref{fig:lag_ene} presents the resulting lag-energy spectrum from the NICER observation in the $0.5-10$ keV band. At high energies ($\gtrsim1$ keV), the time lags exhibit a log-linear trend, indicating that the soft photons are leading the hard photons. Interestingly, a reversal of this trend is observed at low energies ($\leq0.7$ keV), where soft photons appear to lag behind the hard photons.

\begin{figure}
    \includegraphics[width=\columnwidth]{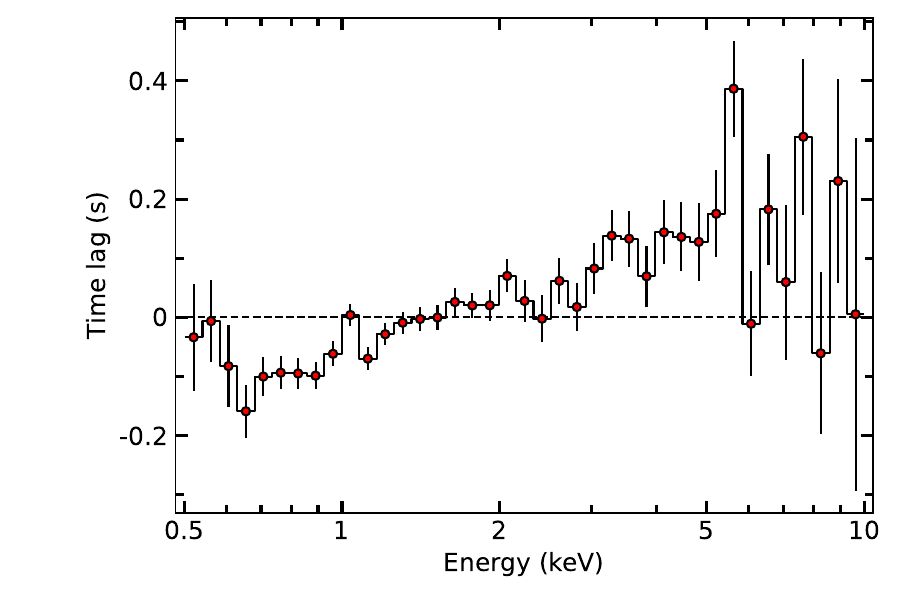}
    \caption{Lag-energy spectrum derived in the $0.5-10$ keV band for the frequency range of $0.02-0.2$ Hz.}
    \label{fig:lag_ene}
\end{figure}

\section{Discussion and Conclusion} \label{sec:discussion}

We have carried out comprehensive broadband spectral and variability studies of the Galactic black hole candidate AT2019wey using quasi-simultaneous Swift/XRT, NICER, and NuSTAR observations during its second outburst in 2022. 

\subsection{Spectral State}

The long-term MAXI light curve shows that the second outburst of AT2019wey, which started in 2022, persisted until early 2025. Comparison of its HID with that of GX~339--4 during its full outburst in 2021 indicates that the 2022 outburst of AT2019wey was a hard-only or failed outburst, with the source remaining in the LHS throughout the entire outburst period without transitioning to the soft state. Similar to this outburst, the previous discovery outburst of the source in 2019 was reported to be a hard-only or failed outburst, during which the source did not transition to the soft state \citep{Yao_2021_ApJ_920_121}. Additionally, hard-only or `failed' outbursts have also been observed in several low-mass BHXRBs such as H~1743--322 \citep{Stiele2021ApJ...914...93S, Sahu2024} and GX~339--4 \citep{Alabarta2021MNRAS.507.5507A, Chand2024ApJ...972...20C}.

\subsection{Comptonizing Regions and Accretion Disk}

Our broadband spectral analysis in the $0.5–78$ keV band, using two different model combinations, indicates that models with only a single Comptonizing region cannot adequately explain the observed spectra. Instead, models that involve two separate Comptonizing regions, along with their corresponding reflection components and thermal emission from the accretion disk, are needed to properly explain the broadband data.

From both of our models, the two Comptonizing regions are found to exhibit distinct physical properties. The first region produces a harder spectrum ($\Gamma \sim 1.6$), with a lower optical depth ($\tau \approx 3$) and a higher electron temperature ($kT_{\rm{e}} \sim 28-370$ keV). In contrast, the second region is relatively cooler, with an electron temperature of $kT_{\rm{e}} \sim 1.2-2.8$ keV, and is characterized by a higher optical depth ($\tau \approx 12$). Furthermore, the hard Comptonizing region is found to dominate the overall flux in both the observations, suggesting that it is responsible for producing the reflection features originating from regions of the accretion disk located away from the ISCO. On the other hand, the soft Comptonizing region contributes minimally to the total observed flux, and our spectral analysis indicates that the influence of this region on the reflection spectrum is relatively minor. A similar accretion geometry has been proposed for MAXI~J1820+070 by \citet{Zdziarski2021ApJ...909L...9Z}, in which the distant, narrow reflection features arise due to the hard Comptonizing region, while the broadening is attributed to the soft Comptonizing region.

The best-fit spectral parameters, including the photon index and electron temperature derived from both models, are consistent with the source being in the LHS during both the observations. The obtained value of the inner disk temperature, $kT_{\rm{in}} \sim0.2$ keV, is well consistent with what is typically observed in the LHS of BHXRBs. From the modeling of the reflection spectrum, we find that the accretion disk is significantly truncated away from the ISCO with {$R_{\rm{in}}\sim 16-56~r_{\rm{g}}$} at the source luminosity of $\sim1.9\%~L_{\rm{Edd}}$, computed assuming the black hole mass and distance to be $10~M_\odot$, and $8$ kpc, respectively. The presence of a significantly truncated accretion disk at relatively high Eddington luminosities ($\gtrsim 2\%$) has been reported in the LHS of other low-mass BHXRBs, such as MAXIJ1820+020 and GX339--4, within the framework of two distinct Comptonizing regions \citep{Zdziarski2021ApJ...909L...9Z, Banerjee2024ApJ...964..189B, Chand2024ApJ...972...20C}. We have also compared the inner accretion disk radius estimated from modeling the disk continuum and reflection component using Model-2. For this, we use the relation $R_{\rm{in}} = \chi k^2 d N^{1/2}_{\rm{diskbb}} \cos^{-1/2}i$, where $\chi \equiv10^4 \rm{cm/1~kpc} \approx 3.24\times 10^{-18}$, $k$ is the color correction factor, $d$ is the distance to the source, $N$ is the \texttt{diskbb} normalization, and $i$ is the disk inclination angle. Assuming $i=30^\circ$ and $k=1.7$, we find the inner disk radius to vary from $18~r_{\rm{g}}$ to $27~r_{\rm{g}}$, considering the uncertainties on the \texttt{diskbb} normalization. These values are in good agreement with those derived from modeling the reflection component. Furthermore, our spectral results indicate that the disk is low to moderately ionized, which is usually expected in the truncated disk scenario. We also notice that keeping the iron abundance fixed at the solar value results in a high density of the disk, which is physically anticipated for BHXRBs \citep{Jiang2019MNRAS.484.1972J}. Earlier versions of the \texttt{relxillCp} model did not allow the disk density to vary freely, instead fixing it at $10^{15}~\rm{cm^{-3}}$, which artificially led to the estimation of super-solar iron abundance \citep{Garcia2018ASPC..515..282G, Zdziarski2022ApJ...928...11Z}.

\subsection{Disk vs Soft Comptonizing region Variability}

The PDSs and CPDSs exhibit nearly identical shapes in both observations (see Figure~\ref{fig:best-fit PDSs}) and are strongly dominated by BLN components, with no evidence for QPOs. The presence of two broad bumps in the PDSs and CPDSs is a characteristic feature in the hard state of BHXRBs, and has been previously observed in the LHS of low-mass BHXRBs, such as GX~339--4 \citep{Nowak1999ApJ...517..355N, Wilkinson2009MNRAS.397..666W, Grinberg2014A&A...565A...1G}. Absence of prominent QPO and strong dominance of aperiodic noise components are also reported in the LHS of AT2019wey during its first outburst by \cite{Yao_2021_ApJ_920_121}. However, the authors reported the detection of a weak type-C QPO at $\sim 2$ Hz as the source transitioned to the HIMS, along with an evolution of the QPO frequency to $\sim 6.5$ Hz with the increase in the disk flux. \cite{Mereminskiy2022A&A...661A..32M} also reported the presence of a strong low-frequency QPO at $\sim55$ mHz in the CPDS derived using NuSTAR observation taken in the LHS during the outburst phase in 2020. 

In the present study, we find that at low frequencies, the power in the $0.5–10$ keV band exceeds that in the $10–30$ keV band. At high frequencies, the power in the soft and hard bands seems to align more closely, although the uncertainties are relatively large. A similar trend is observed in the PDSs of the $0.5–1$ keV and $5–10$ keV bands derived from the NICER data. These results indicate that the accretion disk contributes more prominently to variability on longer timescales than the corona on the shorter timescales. Notably, the long-timescale variability at low frequencies may be influenced by emissions from both the disk and soft Comptonizing region, as suggested by our spectral analysis. However, disentangling the individual variability contributions from these two components is beyond the scope of this paper. Further high-quality observations and frequency-resolved spectroscopy may shed light on the nature of such complex variability in a detailed way. Similar shapes of the PDSs in the LHS of BHXRBs SWIFT~J1753.5--0127 and GX~339--4 were also reported by \cite{Wilkinson2009MNRAS.397..666W}, where the variability power was found to be relatively larger at low frequencies and eventually becomes comparable at high frequencies. The authors suggested that this additional disk variability on long timescales is caused by the instability in the disk.

 The nature of the rms-energy spectra remains consistent across both observations, as illustrated in Figure~\ref{fig:rms_E}. At low energies ($\lesssim3$ keV), where the emission is primarily dominated by either the disk or the soft Comptonizing region, the variability amplitude is high, exceeding $30\%$. On the other hand, the variability of the high-energy ($\gtrsim4$ keV) photons that are mainly coming from the coronal regions remains flat with an amplitude of $\lesssim30\%$. These findings suggest that the disk in the LHS is strongly variable, unlike a stable disk often seen in the HSS, or that the low-energy variability is regulated by the soft Comptonizing region. Such behavior is also consistent with the trends observed in the PDSs and CPDSs. Evidence of decreasing variability with increasing energy in the LHS was also reported for this source by \cite{Yao_2021_ApJ_920_121}, using the observations during its previous outburst in 2019. A similar kind of shape of the rms-energy spectrum was also reported by \cite{Gierlinski2005MNRAS.363.1349G} for the low-mass BHXRB XTE~J1650--500 in the LHS using RXTE observations, where the variability was larger at low energies than at high energies. Interestingly, results from our spectral analysis indicate that the source was in the LHS with a low disk fraction ($\lesssim9\%$), implying that the disk contributes minimally to the total flux. Nevertheless, if such large-amplitude variability at low energies is associated with the disk, it may originate from intrinsic disk fluctuations, likely driven by instabilities within the disk itself.

Detection of hard time lags at low frequencies is commonly attributed to the propagation of mass accretion rate fluctuations from the outer to the inner regions of the accretion disk. This further supports the scenario in which the additional disk variability, observed through the study of the power density spectra (PDS) and rms-energy spectra, arises due to the inward propagation of these fluctuations. In the propagating fluctuation model proposed by \citet{Lyubarskii1997}, variations in the mass accretion rate originate at different radii in the disk and gradually move inward toward the central object. Consequently, these fluctuations first perturb the soft photons emitted from the outer disk regions before influencing the harder photons produced closer to the black hole, leading to a delay of the hard photons relative to the soft ones. An alternative explanation for hard lags is Comptonization delay, wherein high-energy photons are delayed due to scattering in the hot corona. However, the amplitude of lags produced by Comptonization is generally smaller than those caused by propagation effects, and the resulting hard lags do not follow a strictly log-linear trend with photon energy \citep{Uttley2011MNRAS.414L..60U}.

The log-linear behavior of the lag-energy spectrum at $\gtrsim 1$ keV indicates that hard photons lag behind soft ones, and is consistent with the viscous propagation model \citep{Lyubarskii1997, Kotov2001MNRAS.327..799K, Arevalo2006MNRAS.367..801A}. The uncertainties in the time lags at high energy bins are relatively large due to poor data quality. However, at $\lesssim1$ keV, the lag-energy spectrum becomes flat and then takes an inverse turn at $\lesssim0.7$ keV. The flattening of the lag-energy spectrum at $\lesssim1$ keV may be caused by the association of the reverberation lags with the hard lags at low frequencies \citep{DeMarco2015ApJ...814...50D}. Moreover, the turnover suggests that, at $\lesssim0.7$ keV, the disk variability begins to lag behind the powerlaw variability. This switch in the lag behavior is interpreted as thermal reverberation lags, arising due to the thermal reprocessing of the hard coronal X-ray onto the accretion disk, and the delay is caused by the light travel time difference between the direct continuum and the reflected component. It is worth noting that the lag due to thermal reverberation becomes visible at high frequencies, where the hard lags are sufficiently suppressed by the soft lags. However, \citet{DeMarco2015ApJ...814...50D} suggested that reverberation lags may also appear at low frequencies if the power-law component, which dominates at high energies, exhibits substantial aperiodic variability over a wide range of timescales and is thermally reprocessed in the accretion disk. We also find that the energy bands where the thermal reverberation signature is observed are also strongly dominated by the emission from the accretion disk (see Figure~\ref{fig:mod1}). The possible presence of weak reverberation lags in AT2019wey was also suggested by \citet{Wang2022ApJ...930...18W} based on NICER observations obtained in 2020. Additionally, evidence for a thermal reverberation lag with an amplitude of $\sim$60 ms in the lag-energy spectrum, averaged over the 0.1–1 Hz frequency range, was previously reported for the low-mass Galactic BHXRB H~1743--322 \citep{DeMarco2016ApJ...826...70D}. Similar thermal reverberation lags have also been detected in the low-mass BHXRB GX~339--4 \citep[see][]{DeMarco2015ApJ...814...50D}. All the above mentioned spectral and timing results show striking similarity of AT2019wey with Galactic low-mass BHXRBs.

\section{Acknowledgments}
We thank the anonymous reviewer for the useful comments and suggestions that have helped to improve the quality of the paper. This research has made use of archival data of NICER, NuSTAR and Swift (XRT) missions provided by the High Energy Astrophysics Science Archive Research Center (HEASARC), which is a service of the Astrophysics Science Division at NASA/GSFC.  NuSTAR Data analysis was performed using the NuSTAR Data Analysis Software (NuSTARDAS), jointly developed by the ASI Science Data Center (SSDC, Italy) and the California Institute of Technology (USA). PT expresses his sincere thanks to the Inter-University Centre for Astronomy and Astrophysics (IUCAA), Pune, India, for granting support through the IUCAA associateship program. PS is very much grateful to IUCAA, Pune, India, for providing support and local hospitality during her frequent visits to accomplish the final shape of this paper. AAZ acknowledges support from the Polish National Science Center grants 2019/35/B/ST9/03944 and 2023/48/Q/ST9/00138. VKA thanks GH, SAG; DD, PDMSA and Director, URSC for encouragement and continuous support to carry out this research.

\bibliography{reference}{}
\bibliographystyle{aasjournal}

\end{document}